\documentclass[journal,letterpaper]{IEEEtran}
\usepackage[latin1]{inputenc}
\usepackage{amsmath,amsbsy,amssymb}
\usepackage{amsfonts,hyperref}
\usepackage{bbm}
\usepackage{verbatim}
\usepackage{amsthm}
\usepackage{xcolor}

\newcommand{\ketbra}[1]{|#1\rangle\langle #1|}

\newcommand{\eps}{\varepsilon}
\renewcommand{\epsilon}{\varepsilon}
\renewcommand{\phi}{\varphi}
\renewcommand{\theta}{\vartheta}

\newtheorem{theorem}{Theorem}
\newtheorem{lemma}{Lemma}

\newtheorem{remark}{Remark}

\newcommand{\hmax}[1]{{H}_{\max}^{#1}}
\newcommand{\hmin}[1]{{H}_{\min}^{#1}}

\newcommand{\lenc}[1]{\ell_{\rm cmp}^{#1}}
\newcommand{\lext}[1]{\ell_{\rm ext}^{#1}}
\newcommand{\pub}{{\rm pub}}
\newcommand{\prv}{{\rm prv}}

\usepackage{graphicx}

\begin{document}
\author{Joseph M.~Renes and Renato Renner%
\thanks{J.M.\ Renes is with the Institut f\"ur Angewandte Physik, Technische Universit\"at Darmstadt, Hochschulstr. 4a, 64289 Darmstadt, Germany. Email: joe.renes@physik.tu-darmstadt.de.}
\thanks{R.\ Renner is with the Institute for Theoretical Physics, ETH Z\"urich, 8093 Z\"urich, Switzerland. E-mail: renner@phys.ethz.ch.

Copyright (c) 2011 IEEE. Personal use of this material is permitted.  However, permission to use this material for any other purposes must be obtained from the IEEE by sending a request to pubs-permissions@ieee.org.}}

\title{Noisy channel coding via privacy amplification and information reconciliation}

\maketitle 

\begin{abstract}
We show that optimal protocols for noisy channel coding of public or private information over either classical or quantum channels can be directly constructed from two more primitive information-theoretic tools: privacy amplification and information reconciliation, also known as data compression with side information. We do this in the one-shot scenario of structureless resources, and formulate our results in terms of the smooth min- and max-entropy. In the context of classical information theory, this shows that essentially all two-terminal protocols can be reduced to these two primitives, which are in turn governed by the smooth min- and max-entropies, respectively. In the context of quantum information theory, the recently-established duality of these two protocols means essentially all two-terminal protocols can be constructed using just a single primitive. 
\end{abstract}
\maketitle 

\begin{IEEEkeywords} quantum information, channel coding, privacy amplification, information reconciliation, Slepian-Wolf coding, smooth entropies \end{IEEEkeywords}

\IEEEPARstart{O}{ne} of the major trends in information theory, both classical and quantum, is that a small set of proof techniques can be used to construct a wide variety of protocols. Random coding is as ubiquitous as it is useful in classical information theory, and the method of decoupling increasingly plays a similar role in quantum information theory. 
Instead of reusing proofs, a different approach is to reuse the protocols themselves, building up more complicated protocols by combining simpler ones. The goal is to do this in such a way that the inner workings of the parts do not have to be analyzed to ensure the correct functioning of the overall protocol. For instance, joint source-channel coding can be accomplished by simply combining a data compressor with a channel coding scheme~\cite{cover_elements_2006}. In the quantum realm, the ``mother of all'' protocols, a fully-quantum version of the Slepian-Wolf task, can generate a variety of two terminal protocols involving entanglement when combined with teleportation and dense coding~\cite{abeyesinghe_mother_2009}.

In this paper we construct optimal protocols for communication of classical information over noisy channels from two simpler primitives: randomness extraction and information reconciliation, also known as data compression with side information. The construction works for either classical channels or quantum channels explicitly accepting classical inputs, and by replacing randomness extraction with privacy amplification, we directly obtain a protocol for private communication. 
We work in the one-shot scenario of structureless resources, meaning the coding scheme does not rely on repeated uses of a memoryless channel. Rather, the one-shot scenario is considerably broader in approach, encompassing not only channels in the traditional sense of communication (both with and without memory), but also channels as models for the dynamics of a physical system, for which the memoryless assumption would be out of place. 

Besides adopting a new technique to construct the protocols, the resulting capacity expressions are novel  as well. We find that the capacities of a channel for one-shot public~\footnote{We use `public communication' to refer to the usual task of sending classical information, in order to better distinguish it from the case of private communication.} or private communication can be characterized in terms of smooth conditional min- and max-entropies, introduced and characterized in~\cite{renner_smooth_2004,renner_security_2005,tomamichel_duality_2009}. Furthermore, these expressions are shown to be essentially tight, up to small additive terms. 
Appealing to the asymptotic equipartition property (AEP) for smooth entropies~\cite{tomamichel_fully_2009} allows us to quickly recover the usual capacity expressions in the memoryless case, from Shannon's original result on the capacity of the classical channel for public communication~\cite{shannon_mathematical_1948} and the associated capacity for private communication~\cite{wyner_wire-tap_1975,ahlswede_common_1993,preneel_information-theoretic_2000}, to the capacity of a quantum channel for public classical communication (known colloquially as the Holevo-Schumacher-Westmoreland, or HSW, Theorem)~\cite{holevo_capacity_1998,schumacher_sending_1997} as well as for private communication~\cite{devetak_private_2005}. Furthermore, dividing the problem of noisy channel communication into questions of coding and questions of channel properties considerably simplifies the logical arguments and should be of independent pedagogical value.  

One-shot expressions for the capacity of public communication have been derived before. In~\cite{renner_single-serving_2006} the one-shot capacity of a classical channel was characterized in terms of smooth min- and max-entropies, while~\cite{mosonyi_generalized_2009} derives an expression for the capacity of quantum channels in terms of generalized (R\'enyi) relative entropies following a hypothesis-testing approach. These results can be seen as generalizations of earlier (asymptotic) results based on the information spectrum method~\cite{hayashi_general_2003,bowen_quantum_2006}. Very recently,~\cite{wang_one-shot_2010} finds tight bounds on the capacity in terms of a smooth relative entropy quantity again from a hypothesis-testing approach.  Combining the latter results with those here implies a relation between the smooth relative and conditional entropies.

Both of the primitive tasks used here are designed to manipulate ``static'' resources, in the sense that the goal is to transform randomness shared by distant parties into a different form (so that the joint distribution of the values held by the parties is close to a given one). This task should be performed using local operations and a limited amount of communication. In particular, randomness extraction corresponds to the task of generating uniformly-distributed random variables out of non-uniform inputs, while information reconciliation uses classical communication to correlate, or reconcile, a random variable held by one party (Alice) with that held by another (Bob). One can view the classical data transmitted for this latter task as a compression of Alice's random variable, as it can be decompressed by Bob with the help of his random variable (or quantum system).

\begin{figure}[h]
\begin{center}
\includegraphics{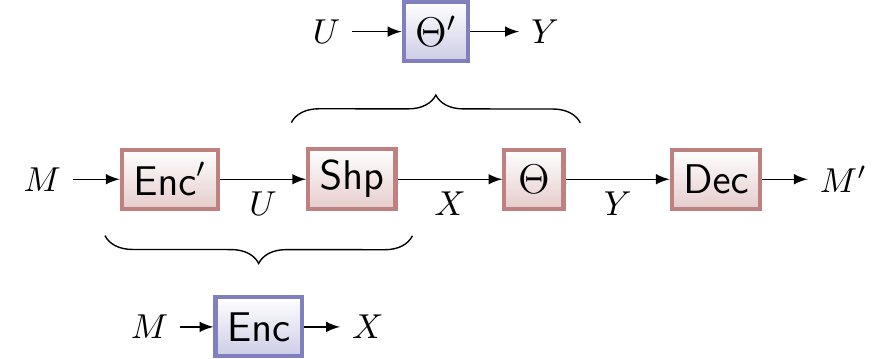}
\end{center}
\caption{\label{fig:diagram}Schematic of using randomness extraction and information reconciliation (data compression with side information) to perform noisy channel communication. 
Messages $m\in M$ are input to the encoder $\mathsf{Enc}'$ and subsequently to the shaper $\mathsf{Shp}$, which is a randomness extractor run in reverse. Then they are then transmitted over the channel ${\Theta}$ to the receiver, who uses the decoder $\mathsf{Dec}$ to construct a guess $m'\in M'$ of the original input. Concatenating the shaper and channel gives a new effective channel ${\Theta}'$, for which an encoder/decoder pair $(\mathsf{Enc}',\mathsf{Dec})$ can be constructed by repurposing a compressor/decompressor pair that operates on the joint input-output $UY$ of the channel. Ultimately, the shaper can instead be regarded as part of the encoder $\mathsf{Enc}$, which is formed by concatenating $\mathsf{Enc}'$ and $\mathsf{Shp}$.
}
\end{figure}

Intuitively, it is plausible that the static information reconciliation protocol could be adapted to enable reliable communication over a noisy channel, a ``dynamic'' resource, in the following manner, depicted schematically in Figure~\ref{fig:diagram}. Assuming uniform distribution of the channel inputs, consider the random variables describing the input $X$ and output $Y$ (where the latter may be quantum in the case of a quantum channel). As described above, information reconciliation enables Bob to reconstruct $X$ from the compressed version of it, $C$, along with his information $Y$. Now suppose Alice and Bob agree on a particular $C=c^*$ in advance for the communication task, in that Alice restricts her channel inputs $X$ to those with compressed output $c^*$. Upon receipt of $Y$, Bob can reconstruct $X$ by simply reusing the decompressor of the information reconciliation protocol. 
In this way, each information reconciliation protocol defines a channel coding scheme: every value $C=c$ specifies a channel code consisting of all the possible inputs which compress to that value $c$. Since we assumed uniform distribution of the channel inputs, this coding scheme is generally not optimal. However, by running a randomness extractor backwards we can create the optimal input distribution from a uniform one, circumventing this problem.  This is similar to a method used by Gallager~\cite{gallager_information_1968} and later expanded in~\cite{ma_matched_2002,soriaga_distribution_2003}.

The remainder of the paper is devoted to making this intuition rigorous. We begin in the next section by formally specifying the problem and stating the results in Theorem~\ref{thm:capacity}. We then move immediately to the proof of the direct part, achievability, in Section~\ref{sec:achieve} and the converse in Section~\ref{sec:converse}. In Section~\ref{sec:asympt} we show how the usual results may be quickly recovered for the case of very many uses of a memoryless channel. We conclude in Section~\ref{sec:concl} by discussing applications of this result and its relation to other work.

\section{Definitions and Results}

We work directly with a classical-quantum channel $\Theta$ taking input classical symbols $x\in\mathcal{X}$ to output quantum states $\theta_x^Y\in\mathcal{S}(\mathcal{H}^Y)\equiv\mathcal{Y}$. To recover the case of a classical channel, one can simply require that the output states $\theta_x^Y$ be simultaneously diagonalizable.

We note that the restriction to classical channel inputs can be made without loss of generality as we are interested in transmitting classical information, either publicly or privately. In terms of a physical channel accepting quantum inputs, this just amounts to fixing the quantum states to be input for given classical value $x$; here this choice is effectively part of the channel (this is possible since, in the one-shot treatment adopted here, the channel is only used once). On the other hand, one may regard this choice as part of the encoder, and the only necessary modification of the expression for the capacity (see Theorem~\ref{thm:capacity} below) would be to include an optimization over this choice.

An $(n,\eps)$-coding scheme for a classical-quantum channel consists of an encoder $\mathsf{Enc}:\mathcal{M}\rightarrow \mathcal{X}$ taking classical messages $m\in \mathcal{M}$ to channel inputs and a decoder $\mathsf{Dec}:\mathcal{Y}\rightarrow \mathcal{M}$ taking channel outputs to guesses of the input messages, for which $n=\log_2|\mathcal{M}|$ and 
\begin{align}
p_{\rm error}(m)&\equiv{\rm Pr}[m\neq \mathsf{Dec}\circ\Theta\circ\mathsf{Enc}(m)]\leq \eps
\end{align}
for all $m\in \mathcal{M}$. In addition, if $\Theta$ outputs a bipartite state ${\theta}_x^{YZ}$, of which Bob receives the $Y$ subsystem and an eavesdropper Eve the $Z$ subsystem living in $\mathcal{S}(\mathcal{H}^Z)\equiv\mathcal{Z}$, then an $(n,\eps)$-private coding scheme is an encoder-decoder pair as above, with the additional requirement that every message $m$ be approximately unknown to the eavesdropper: 
\begin{align}
\label{eq:secdef}
p_{\rm secret}(m)&\equiv\tfrac{1}{2}
\big\|{\theta}_{\mathsf{Enc}(m)}^Z-{\theta}^Z\big\|_1\leq \eps,
\end{align}
where ${\theta}^Z=\tfrac{1}{2^n}\sum_m \theta^Z_{\mathsf{Enc}(m)}$\footnote{In fact, $\theta^Z$ need not be the average state, but can be arbitrary.}.

It is useful to think of the message as being a random variable $M$, taking values in $\mathcal{M}$ according to the probability distribution $P_M$. Then the message transmission process is encapsulated by the following sequence of random variables (Markov chain). We use a prime to denote a random variable or quantum system which is meant to be nearly identical to the unprimed version; in the present context we would like the output $M'$ to be essentially equal to the input $M$.
\def\arrlen{1.0cm}
\begin{align}
M\,\underset{\mathsf{Enc}}{\xrightarrow{\hspace*{\arrlen}}}\,X\,\underset{\Theta}{\xrightarrow{\hspace*{\arrlen}}}\,Y\,\underset{\mathsf{Dec}}{\xrightarrow{\hspace*{\arrlen}}}\,M'
\end{align}

Given a choice of $\eps$, the capacity for public communication (usually just referred to as the classical capacity) $\mathcal{C}^\eps_\pub(\Theta)$ of the channel is simply $\log_2 n$ for the largest $n$ in an $(n,\eps)$ coding scheme. The private capacity $\mathcal{C}^\eps_\prv(\Theta)$ is defined similarly using private coding schemes. Here we prove the following upper and lower bounds on these capacities in terms of the smooth min- and max-entropies, which are defined in the appendix.
\begin{theorem}[Capacities of Classical-Quantum Channels] \label{thm:capacity}
$\phantom{.}$\\For all $\eps>0$, \begin{align}
 C_\pub^{\eps}(\Theta)\geq \max_{P_X}\Big[&\hmin{\eps/8}(X)-\hmax{\eps/8}(X|Y)\nonumber\\&-4\log\tfrac{1}{\eps}-16\Big],\\ 
C_\pub^{\eps}(\Theta)\leq \max_{P_X}\Big[&\hmin{}(X)-\hmax{\sqrt{2\eps}}(X|Y)\Big].
\end{align}
For $T\rightarrow X\rightarrow (Y,Z)$ a Markov chain,
\begin{align}
C_\prv^{8\eps}(\Theta)\geq\max_{P_T, T\rightarrow X}\Big[&\hmin{\eps/8}(T|Z)-\hmax{\eps/8}(T|Y)\nonumber\\
&-4\log\tfrac{1}{\eps}-16\Big],\\
C_\prv^{\eps}(\Theta)\leq \max_{P_T, T\rightarrow X}\Big[&\hmin{\sqrt{2\eps}}(T|Z)-\hmax{\sqrt{2\eps}}(T|Y)\Big].
\end{align}
\end{theorem}

\section{Achievability}
\label{sec:achieve}
The proof of the direct parts proceeds in three steps, successively building up to a construction of an encoder and decoder. The first step is to show that in doing this, we only need to worry about the average transmission error and secrecy of the communication scheme, assuming the inputs are uniformly distributed. Then, we show that protocols for information reconciliation 
can be adapted to the channel coding scenario, when the input to the channel is uniformly distributed (not just the messages themselves). Finally, we show how to mimic any particular channel input distribution from a uniform distribution by using randomness extraction, and how to mimic an input distribution so that the eavesdropper learns nothing about the message using privacy amplification.

\def\boxsep{1.85}

\subsection{Average Case Coding Implies Worst Case Coding}
We start by observing that constructing an encoder/decoder pair with low \emph{average} error probability on the receiver's end and low average trace distance of eavesdropper outputs suffices to construct an encoder/decoder pair with low error probability and secrecy parameter in the worst case. 
\begin{lemma}[Average Case to Worst Case Error]
\label{lem:avetoworst}
Given a channel $\Theta:X\rightarrow Y$ and an encoder/decoder pair $\mathsf{Enc}:M\rightarrow X$, $\mathsf{Dec}:Y\rightarrow M'$ such that $\frac{1}{|\mathcal{M}|}\sum_{m\in\mathcal{M}} p_{\rm error}(m)\leq \tfrac{\eps}{4}$ and $\frac{1}{|\mathcal{M}|}\sum_{m\in\mathcal{M}} p_{\rm secret}(m)\leq \frac{\eps}{4}$, then 
there exists an encoder/decoder pair for a subset $\mathcal{M^*}\subset \mathcal{M}$ of size at least $|\mathcal{M}|/2$ such that
$p_{\rm error}(m^*)\leq \eps$ and $p_{\rm secret}(m^*)\leq \eps$ for all $m^*\in\mathcal{M}^*$.
\end{lemma}
\begin{IEEEproof}
By the Markov inequality, a fraction at most one-quarter of $m\in\mathcal{M}$ have $p_{\rm error}(m)\geq \eps$. Similarly, for at most one-quarter does $p_{\rm secret}(m)\geq \eps$. Thus, there is a subset $\mathcal{M}^*$ of half the $m\in \mathcal M$ for which neither statement is true. Restricting the input of $\mathsf{Enc}$ to ${\mathcal M}^*$ gives the new encoder. The new decoder is given by altering the old decoder so that outputs $m\notin{\mathcal M}^*$ are mapped at random to $m\in{\mathcal M}^*$. 
\end{IEEEproof}
\begin{remark} 
\label{rem:avetoworst}
If we only require that $p_{\rm error}(m)\leq \eps$, then $\frac{1}{|\mathcal{M}|}\sum_{m\in\mathcal{M}} p_{\rm error}(m)\leq \frac{\eps}{2}$ suffices.
\end{remark}
\subsection{Channel Coding From Information Reconciliation For Uniformly-Distributed Inputs}
Now we show that an information reconciliation protocol can be adapted to channel coding, at least when the input to the channel is uniformly or nearly uniformly distributed. We do this explicitly for the case of linear compression functions and subsequently remark how it can be made more general.

First we need to specify classical-quantum information reconciliation protocols more precisely. Given a classical random variable $X$ and a quantum system $Y$ jointly described by the classical-quantum state $\psi^{XY}=\sum_{x\in\mathcal X}p_x \ketbra{x}^X\otimes \vartheta_x^Y$, an $\eps$-good information reconciliation protocol consists of a compression map $\mathsf{Cmp}:\mathcal{X}\rightarrow\mathcal{C}$ taking $X$ to another classical random variable $C$ and a decompression map $\mathsf{Dcp}: (\mathcal C,\mathcal Y)\rightarrow \mathcal X$ taking $C$ and states in the system $Y$ to elements $X'$ of the input alphabet $\mathcal X$ such that the error probablility $p_{\rm error}=\sum_x p_x {\rm Pr}[x\neq \mathsf{Dcp}(\mathsf{Cmp}(x),\vartheta_x^Y)]\leq \eps$. If the alphabet $\mathcal{X}$ forms a linear space, the compression map could be linear, and one speaks of a \emph{linear compressor}.

\begin{lemma}[Channel Coding from Information Reconciliation of Uniform Inputs]
\label{lem:csicoding}
Given a cq channel $\Theta:\mathcal U\rightarrow \mathcal{Y}$ from uniformly-distributed inputs $U$ to arbitrary outputs $Y$, suppose the linear compressor and arbitrary decompressor pair $\mathsf{Cmp}/\mathsf{Dcp}$ form an $\eps$-good information reconciliation protocol for the combined input and output $UY$. Then there exists a linear encoder $\mathsf{Enc}:\mathcal M\rightarrow \mathcal U$ and a decoder $\mathsf{Dec}:\mathcal Y\rightarrow \mathcal M$ for $\Theta$ such that the error probability of transmitting a uniformly-distributed message $M$ of size $|\mathcal{M}|= |\mathcal U|/|\mathcal C|$ is also less than $\eps$.
\end{lemma}
\begin{IEEEproof}
Start by defining $p_{\rm error}(u)={\rm Pr}[u\neq\mathsf{Dcp}(\mathsf{Cmp}(u),\theta_u)]$. Then the information reconciliation error probability can be formulated as
\begin{align}
p_{\rm error} &= \frac{1}{|\mathcal U|}\sum p_{\rm error}(u) = \frac{1}{|\mathcal C|}\sum_c\frac{|\mathcal C|}{|\mathcal U|}\!\sum_{u:\mathsf{Cmp}(u)=c}p_{\rm error}(u)\nonumber\\
&=\frac{1}{|\mathcal C|}\sum_c \langle p_{\rm error}(u)\rangle_{P_{U|C=c}},
\end{align}
where $\langle X\rangle_{P_X}$ denotes the average of $X$ using the distribution $P_X$ and $P_U$ is the uniform distribution. In other words, the average error probability is the average over outputs $c$ of the average error probability of inputs $u$ consistent with a given output. 
In this expression we have split the summation over $u$ to first a summation over the values of $c$ and then for each of these a summation over the $u$ for which $\mathsf{Cmp}(u)=c$. In so doing, we have used the fact that there are $|\mathcal U|/|\mathcal C|$ preimages for each $c$, which follows from Lemma~\ref{lem:linearpreimage} (see appendix). 

Choosing the value of $C=c^*$ with the lowest error probability $\langle p_{\rm error}(u)\rangle_{P_{U|C=c^*}}$ enables us to define an encoder and decoder from the compressor and decompressor restricted to this value. The encoder simply maps $m\in \mathcal M$ to those $u\in \mathcal U$ for which $\mathsf{Cmp}(u)=c^*$ in some fixed order, say lexicographic order. By linearity of the compressor, $|\mathcal M|=|\mathcal U|/|\mathcal C|$. The decoder is then defined by taking the output of the decompressor and then applying the inverse of the encoding map to the result $u'$ or outputting a random $m\in \mathcal{M}$ when $\mathsf{Cmp}(u')\neq c^*$.

The error probability for the encoder/decoder combination for uniformly distributed messages $M$ is exactly the same as the error probability for the compressor/decompressor combination, which must be lower than the average by construction. 
\end{IEEEproof}

\begin{remark}
\label{rem:csiepscoding}
If the compressor/decompressor pair has error probability $\eps_2$ when acting on a nearly uniform input $U'$ satisfying $\tfrac 12\left\|U-U'\right\|_1\leq \eps_1$, then applying the corresponding encoder/decoder to a uniform input gives an error probability of at most $\eps_1+\eps_2$ by the triangle inequality. 
\end{remark}

\begin{remark} \label{rem:generalcomp} By using Lemma~\ref{lem:arbpreimage} instead of Lemma~\ref{lem:linearpreimage} (see appendix), the restriction to linear compression functions can be removed at the cost of reducing the number of messages by a factor $\eps$ and an additional failure probablity $\eps$.  \end{remark}

\subsection{Distribution Shaping}
Finally, we need to remove the restriction of uniform inputs to the channel. This is done by combining the channel with a \emph{distribution shaper}, which is a means of mapping a uniform distribution to a chosen distribution. By running the distribution shaper and then the channel, we obtain a virtual channel which acts again on a (roughly) uniformly distributed input.
The distribution shaper can be constructed using a randomness extractor, as follows.

Suppose that $\mathsf{Ext}:X\rightarrow U'$ is a function which produces an $\eps$-good approximation of a uniformly distributed random variable $U$ from an input $X$ distributed according to $P_X$, in the sense that $\tfrac 12\left\|U-U'\right\|_1\leq \eps$. The extractor defines a joint distribution $P_{XU'}$, and with this we can define a function $\mathsf{Shp}:(U',R)\rightarrow X$ which is in some sense the inverse of $\mathsf{Ext}$. Here $R$ is some additional randomness, and $\mathsf{Shp}$ is defined by using $R$ to select an $x\in X$ from the distribution $P_{X|U'=u'}$ given the input value $U'=u'$. Thus, the output of the shaper is again $X$. Shapers constructed in this manner will be called $\eps$-shapers. Moreover, if the extractor performs privacy amplification of $X$ against some $Z$ generated from $X$, then the shaper replicates $X$ while hiding $U'$ from the eavesdropper. This follows because the conditional states relevant to the eavesdropper are the same in both cases. 

It may seem strange to additionally require a source of randomness for this purpose, and ideally we would like all the randomness needed to generate $X$ to be contained in $U'$. However, the mapping that takes a general $X$ to a nearly-uniform distribution $U'$ may  map two values $x$ to the same $u'$. When that $u'$ is input to $\mathsf{Shp}$, some randomness is needed to reverse the mapping.

\subsection{Putting it all together}

Now we can combine these three pieces to establish the direct part of Theorem~\ref{thm:capacity}. We do this first for the private capacity and then make some modifications to obtain the lower bound on the classical capacity. The latter can be obtained as a special case of the former by assuming the channel does not leak anything to an adversary, i.e.\ $Z$ is trivial, but the additional modifications will improve the constants in the bound. 

For a given channel $\Theta:X\rightarrow Y$ and input distribution $P_X$, we can define a new channel $\Theta':U'\rightarrow Y$ (with output states $\theta^{'Y}_{u'}$) by concatenating an $\eps_1$-shaper $\mathsf{Shp}$ that generates $X$, built from a privacy amplification extractor, with $\Theta$ and regarding $R$ as part of the channel. Next, following Remark~\ref{rem:csiepscoding} we construct an encoder/decoder pair $\mathsf{Enc}'$/$\mathsf{Dec}'$ from an $\eps_2$-good compressor/decompressor for $\psi^{U'Y}=\sum_{u'\in \mathcal U}p_{u'} \ketbra{u'}^{U'}\otimes\theta^{'Y}_{u'}$, where $p_{u'}$ is the distribution of the input $U'$ to the shaper $\mathsf{Shp}$. When input with a uniformly distributed $U$, the error probability averaged over codewords and choices of code is at most $\eps_1+\eps_2$, while the average leakage to the eavesdropper is at most $2\eps_1$. For simplicity, define $\eps\equiv 4\max(\eps_1,\eps_2)$. By the Markov inequality, at least three-quarters of the code choices have an average codeword error rate below $4(\eps_1+\eps_2)<2\eps$. By the same reasoning, at least three-quarters of the code choices have an average $p_{\rm secret}$ less than $8\eps_1<2\eps$. Therefore, at least half have both properties.

Regarding the shaper as part of the encoder instead of part of the channel, we can define $\mathsf{Enc}=\mathsf{Shp}\circ\mathsf{Enc}'$. 
Applying Lemma~\ref{lem:avetoworst}, we can then make the further adjustments to $\mathsf{Enc}$ and $\mathsf{Dec}$ to simultaneously achieve a worst-case error of $8\eps$ and worse-case leakage $8\eps$. 

Finally, we can count how many messages can be reliably sent using the constructed encoder and decoder. From Lemma~\ref{lem:csicoding}, we have $n\geq \log|\mathcal U|-\log|\mathcal C|-1$. Inserting the known results for privacy amplification and data compression, Theorems~\ref{thm:pa} and~\ref{thm:csi} in the appendix,
this becomes 
\begin{align}
\label{eq:capacitycount}
n\geq \hmin{\eps_{11}}(X|Z)-2\log\tfrac{1}{\eps_{12}}-\hmax{\eps_{21}}(U'|Y)-2\log\tfrac{1}{\eps_{22}}-4,
\end{align}
where $\eps_1=\eps_{11}+\eps_{12}$ and $\eps_2=\eps_{21}+\eps_{22}$. Again for simplicity, let $\eps_{jk}=\eps/8$. Because $U'$ is a function of $X$ via the extractor, the max-entropy cannot increase when replacing $X$ by $U'$. Since we are free to choose any $P_X$ in this argument we therefore have
\begin{align}
n\geq \max_{P_X}\Big[&\hmin{\eps/8}(X|Z)-\hmax{\eps/8}(X|Y)-4\log\tfrac{1}{\eps}-16\Big].
\end{align}
To complete the argument for the private capacity, note that Alice could precede the channel with another mapping from $T$ to $X$, which she is free to optimize. Regarding this as part of the original channel in the above argument then leads to the desired result. 

The direct part for the channel capacity follows by making a few small modifications. First, the Markov inequality is no longer needed to ensure the two conditions of private communication are satisfied. Here there is only one, and certainly there exists an encoding with average codeword error probability less than the average over codes and codewords, $\eps_1+\eps_2$. We then only require Remark~\ref{rem:avetoworst} rather than Lemma~\ref{lem:avetoworst} to move to the worst-case error $2(\eps_1+\eps_2)$ over codewords. Finally, though in principle it is also possible for Alice to precede the channel with a $T\rightarrow X$ mapping, we shall see in the converse that this is not necessary. Note also that in this context the encoder can dispense with the randomness needed to properly simulate $X$ and just fix a particular value of the output, for instance the $x$ with the largest $P_{X=x|C=c^*}$.

\section{Converse}
\label{sec:converse}
We first prove the converse for the private capacity and then modify the argument to establish the converse for the classical capacity. 
Given an $(n,\eps)$-private coding scheme, the two requirements of the output made by the definition imply that $\hmax{\sqrt{2\eps}}(M|M')\leq 0$ and $\hmin{\sqrt{2\eps}}(M|Z)\geq n$, the former for any distribution of messages and the latter for the uniform distribution. The former follows because the trace distance of the pair $(M,M')$ to $(M,M)$ is less than $\eps$ and therefore $(M,M)$ is in the $\sqrt{2\eps}$-neighborhood of $(M,M')$ (see the appendix; the square root is a consequence of the conversion of the trace to purification distance). But $\hmax{}(M|M)=0$ and thus $\hmax{\sqrt{2\eps}}(M|M')\leq 0$. The latter follows because again the ideal output, in which $Z$ is independent of the uniformly-distributed $M$ and therefore satisfies $\hmin{}(M|Z)\geq n$, is in the $\sqrt{2\eps}$-neighborhood of the actual pair $(M,Z)$.
Additionally, by the data processing inequality~\cite{tomamichel_duality_2009}, $\hmax{\sqrt{2\eps}}(M|M')\geq\hmax{\sqrt{2\eps}}(M|Y)$ since the decoder generates the guess $M'$ from $Y$.

Defining $\bar{P}_M$ to be the uniform distribution, we have 
\begin{align*}
\max_{P_M,M\rightarrow X}&\left[\hmin{\sqrt{2\eps}}(M|Z)-\hmax{\sqrt{2\eps}}(M|Y)\right]\\&\geq \max_{M\rightarrow X}\left[\hmin{\sqrt{2\eps}}(M|Z)_{\bar{P}_M}-\hmax{\sqrt{2\eps}}(M|Y)_{\bar{P}_M}\right]\\& \geq \max_{M\rightarrow X}\left[\hmin{\sqrt{2\eps}}(M|Z)_{\bar{P}_M}-\hmax{\sqrt{2\eps}}(M|M')_{\bar{P}_M}\right]\\
&\geq n,
\end{align*}
which is the form we set out to prove.

For the converse of the classical capacity, observe that the encoding function $\mathsf{Enc}$ is without loss of generality deterministic and injective. It might as well be deterministic, since if it used randomness, we could make it deterministic by fixing the randomness to that value with the least probability of error, which cannot be worse than the average case. Moreover, for this deterministic choice, $\mathsf{Enc}$ must be injective, since a collision of two inputs having the same codeword necessarily implies an error. Now, using the injectivity of $\mathsf{Enc}$ we can define a distribution $\bar{P}_X$ given a distribution over $M$ by simply taking the distribution of $M$ on its image in $X$ and zero otherwise. Choosing the uniform distribution over $M$ and observing that $\hmin{}(M)=n$ when $M$ is uniformly distributed, we obtain 
\begin{align}
\max_{P_X}&\left[\hmin{}(X)-\hmax{\sqrt{2\eps}}(X|Y)\right]\\&\geq \left[\hmin{}(X)_{\bar{P}_X}-\hmax{\sqrt{2\eps}}(X|M')_{\bar{P}_X}\right]\\
&= \left[\hmin{}(M)-\hmax{\sqrt{2\eps}}(M|M')\right]\\&\geq n.
\end{align}

\section{Asymptotic Analysis} \label{sec:asympt}
In the asymptotic limit of $n\rightarrow\infty$ uses of a memoryless channel we recover the known results on the \emph{rate} of public or private communication, where the rate of private communication is defined by 
\begin{align}
R_\prv(\Theta)=\lim_{\eps\rightarrow 0}\lim_{n \rightarrow \infty}\frac{\mathcal{C}^{\eps}_\prv(\Theta^{\otimes n})}{n},
\end{align}
and the rate of public communication is defined similarly. In general, the rates take the rather ungainly form
\begin{align}
\label{eq:classrate}
R_\pub(\Theta)&=\lim_{\ell\rightarrow\infty}\frac{1}{\ell}\max_{P_{X^\ell}}\left[H(X^\ell)-H(X^\ell|Y^{\otimes \ell})\right],\\
\label{eq:privrate}
R_\prv(\Theta)&=\lim_{\ell\rightarrow\infty}\frac{1}{\ell}\max_{P_{T},T\rightarrow X^\ell}\left[H(T|Z^{\otimes \ell})-H(T|Y^{\otimes \ell})\right].
\end{align}
Here $X^n$ refers to a classical random variable on $\mathcal{X}^n$, while $Y^{\otimes n}$ refers to the $n$-fold tensor product of the Hilbert space $\mathcal{H}^Y$ and similarly for $Z$. The rate for public communication over quantum channels is known as the HSW theorem, after Holevo~\cite{holevo_capacity_1998} and Schumacher and Westmoreland~\cite{schumacher_sending_1997}. The private rate was proven by Devetak~\cite{devetak_private_2005}.

For the special case of classical channel outputs, i.e.\ $\theta^Y_x$ (and separately $\widetilde{\theta}^Z_x$) are all simultaneously diagonalizeable, these reduce to the familiar and simpler form
\begin{align}
R_\pub(\Theta_{\rm cl})&=\max_{P_X}\left[H(X)-H(X|Y)\right],\label{eq:clasyrate}\\
R_\prv(\Theta_{\rm cl})&=\max_{P_T,T\rightarrow X}\left[H(T|Z)-H(T|Y)\right].\label{eq:prasyrate}
\end{align}
The classical rate is Shannon's original noisy channel coding theorem~\cite{shannon_mathematical_1948}. The private rate was first established by Wyner in the specific setting of the wire-tap channel~\cite{wyner_wire-tap_1975}, later expanded to arbitrary channels by Ahlswede and Csiszar~\cite{ahlswede_common_1993}, and strengthened to the stronger form of security used here (cf.\ Eq.~\ref{eq:secdef}) by Maurer and Wolf~\cite{preneel_information-theoretic_2000}.

The proof proceeds in two steps. First we show that such rates are possible using the lower bound on the capacity and applying the asymptotic equipartition property (AEP) of the conditional min- and max-entropies. Then we show that the rates cannot be exceeded by making use of the upper bound on the capacity and bounds on the conditional min- and max-entropy in terms of the conditional von Neumann entropy. Since the case of public communication follows from that of private communication, we only give the argument for the latter.

For the direct part, we begin with the lower bound on the capacity from Theorem~\ref{thm:capacity}. For $m$ uses of the channel $\Theta$, this becomes
\begin{align}
\mathcal{C}^{8\eps}_\prv(\Theta^{\otimes m})\geq \max_{P_{T},T\rightarrow X^m}\!\Big[&\hmin{\eps/8}(T|Z^{\otimes m})-\hmax{\eps/8}(T|Y^{\otimes m})\nonumber\\
&-4\log\tfrac{1}{\eps}-16\Big].
\end{align}
Since this is a lower bound, we're free to choose $T$ as we like. We choose $T=T^m$ to be i.i.d., each instance $T_i$ separately generating the likewise i.i.d.\ $X_i$ via some  fixed map $T\rightarrow X$. Now we make use of the AEP, which states~\cite{tomamichel_fully_2009}
\begin{align}
\lim_{\eps\rightarrow 0}\lim_{m\rightarrow \infty}\hmin{\eps}(X^m|Z^{\otimes m})=mH(X|Z),
\end{align}
and similarly for the conditional max-entropy. We then obtain
\begin{align}
\lim_{\eps\rightarrow 0}\lim_{m\rightarrow \infty}\frac{1}{m}\mathcal{C}^{8\eps}_\prv(\Theta^{\otimes m})\geq \max_{P_{T},T\rightarrow X}\left[H(T|Z)-H(T|Y)\right].
\end{align}
Finally, we let $n=\ell m$, and replay the above argument using the superchannel $\Theta^{\otimes \ell}$ ($\ell$ independent uses of the channel) to obtain the desired result.  

Note that, in contrast to the standard proof technique, here we only need to make a statement about the entropies in the capacity formula, a statement provided by the AEP. Importantly, typical sequences, type classes, or the like play no role in the protocol itself, but could be used to establish the AEP. Such methods are not necessary; indeed, the approach of~\cite{tomamichel_fully_2009} is based on properties of R\'enyi entropies.

To complete the argument, we consider the upper bound on the capacity, and use the bounds on the conditional min- and max-entropies from Lemma~\ref{lem:entropybounds}, replacing dim$(A)$ with $|\mathcal{X}|$. Now the upper bound on the private capacity becomes
\begin{align}
\mathcal{C}^\eps_\prv(\Theta^{\otimes n})\leq \!\max_{P_{T},T\rightarrow X^n}\!\Big[&H(T|Z^{\otimes n})-H(T|Y^{\otimes n})\nonumber\\
&+16n\sqrt{2\eps}\log|\mathcal{X}|+4h_2(2\sqrt{2\eps})\Big].
\end{align}
Dividing through by $n$ and taking the limits $n\rightarrow\infty$ and $\eps\rightarrow 0$ (whose order is now irrelevant) yields the desired result (replacing $n$ by $\ell$).

When the channel has purely classical outputs, the limit involving $\ell\rightarrow \infty$ (called regularization) can be removed. For private communication we show this explicitly in Lemma~\ref{lem:singlelett}, which recovers Eq.~\ref{eq:clasyrate}; for public communication (Eq.~\ref{eq:prasyrate}) see, e.g.\ Theorem 4.2.1 in~\cite{gallager_information_1968}. Should the channel produce quantum outputs, regularization is known to be necessary in both cases, private~\cite{smith_structured_2008} and public~\cite{hastings_superadditivity_2009}. 

\begin{figure}
\def\colgap{2.5}
\def\rowgap{1}
\begin{center}
\includegraphics{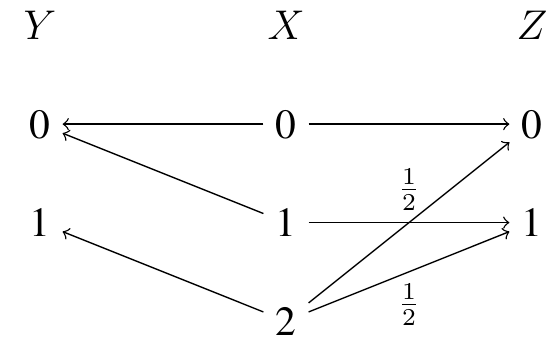}
\end{center}
\caption{
\label{fig:example}
Channel demonstrating the need for randomness by the encoder to achieve the private communication capacity. Unmarked arrows denote deterministic maps; otherwise the probability of a transition is marked.}
\end{figure}

Finally, we note that the optimization over maps $T\rightarrow X$ is generally necessary to achieve the optimal rate of private communication, by means of the following example. Suppose $\Theta$ is a purely classical channel defined in Fig.~\ref{fig:example}. To send private messages to $Y$, clearly one can encode 0 as $X=0$ or $X=1$ randomly and 1 as $X=2$. The message can be unambiguously determined from $Y$ no matter the encoding, but $Z$ will be completely random for either input, and so this encoding scheme achieves a rate of 1 bit. Eschewing random encoding, the maximum rate of private communication is $\max_{P_X}\left[H(X|Z)-H(X|Y)\right]$. Due to the structure of the $X\rightarrow Z$ map, the maximum of the first term is one, and this can only occur when 0 and 1 occur with equal probability. But this implies the second term is nonzero, meaning the overall rate is less than one.

\section{Conclusions} \label{sec:concl}
Rather than reusing the proof techniques involved in understanding more basic information processing protocols such as privacy amplification and information reconciliation, here we have shown how to construct channel coding protocols from these protocols themselves. Moreover, if the underlying protocols are optimal, then so are the channel coding protocols. This provides an appealing conceptual framework for two-terminal problems in information theory in which one successively builds up to more complicated protocols using simpler elements whose internal workings are not relevant for the present task. Moreover, it is also appealing to see that the two basic primitives are characterized in terms of the two basic entropic quantities, smooth min- and max-entropy, and that these quantities enter the capacity expressions in a way which reflects the protocol construction.

In the setting of quantum information theory these entropies are dual~\cite{tomamichel_duality_2009}, as are the two primitives~\cite{renes_duality_2010}, meaning only one primitive is needed to construct more and more complicated protocols. As an example, instead of appealing to Theorem~\ref{thm:csi} for the compressor/decompressor pair needed to establish Eq.~\ref{eq:capacitycount} in Theorem~\ref{thm:capacity}, we may rely on 
\begin{enumerate}
\item Theorem~\ref{thm:pa},
\item the duality of privacy amplification and information reconciliation as shown in~\cite{renes_duality_2010}, and
\item a new form of the uncertainty principle derived in~\cite{tomamichel_uncertainty_2010}.
\end{enumerate}
 Specifically, in the proof of Theorem~\ref{thm:capacity} we require a linear compressor/decompressor pair operating on the classical-quantum state of $U'Y$, classical in say the $U'$ basis (in an abuse of notation). By the duality in~\cite{renes_duality_2010}, such a pair with error probabilty $\sqrt{2\eps}$ can be constructed from $\eps$-good linear privacy amplification of the conjugate basis $\widetilde{U}'$~\footnote{The cq nature of the state ensures that we satisfy condition (b) of Theorem 4 in~\cite{renes_duality_2010}.}, and the size of the compressed output $\mathcal{C}$ in the optimal case is given by $\log_2 |\mathcal{C}|=\log_2 |\mathcal{U}|-\hmin{\eps_1}(\widetilde{U}'|R)+2\log\frac{1}{\eps_2}-1$, where $R$ is the purification of the original system $XY$ and $\eps=\eps_1+\eps_2$. But from the uncertainty principle of~\cite{tomamichel_uncertainty_2010} we have 
\begin{align} 
\hmin{\eps}(\widetilde{U}'|R)+\hmax{\eps}(U'|Y)\geq \log |\mathcal{U}|.
\end{align}
Therefore $\log_2|\mathcal{C}|\leq\hmax{\eps}(U'|Y)+O(\log\frac{1}{\eps})$, and we recover Eq.~\ref{eq:capacitycount} up to  $O(\log\frac{1}{\eps})$ terms.\footnote{$\log_2|\mathcal{C}|$ also cannot be substantially smaller, by the lower bound of Theorem~\ref{thm:csi}), and the capacity as calculated here is certainly no smaller than that of Eq.~\ref{eq:capacitycount} (up to $O(\log\frac{1}{\eps})$ terms).} Thus we have constructed all-decoupling proofs 
of the public and private capacities of a quantum channel, in the sense that establishling the capacity now does not rely on directly constructing a decoder for the receiver, as in the proof of Theorem~\ref{thm:csi}, but rather on decoupling the purifying system, as in the proof of Theorem~\ref{thm:pa}.\footnote{Observe that we could not have used a similar form of the uncertainty principle derived in~\cite{renes_duality_2010} for the present purpose, as it subtly relies on the decoder construction we are trying to avoid.} In the asymptotic limit of many independent uses of the channel, we then recover the familiar HSW~\cite{holevo_capacity_1998,schumacher_sending_1997} and private capacity~\cite{devetak_private_2005} results.  This derivation of the classical capacity of a quantum channel can thus be seen as the classical-quantum analogue of~\cite{hayden_decoupling_2008}, where the quantum capacity of a quantum channel is derived using a decoupling approach. As with the main proof of Theorem~\ref{thm:capacity}, the decoupling procedure outlined above also suffices to derive Shannon's result on the public capacity of classical channels~\cite{shannon_mathematical_1948}, as well as the associated privacy capacity results~\cite{wyner_wire-tap_1975,ahlswede_common_1993,preneel_information-theoretic_2000}, simply by treating the classical channel in the formalism of quantum information theory.

\section*{Acknowledgments}
JMR acknowledges the support of CASED (www.cased.de). RR acknowledges support from the Swiss National Science Foundation (grant No.~200021-119868) and the ERC (grant No. 258932).

\appendix
\section{Smooth Min- and Max-Entropies}
The conditional max-entropy for a state $\rho^{AB}$ is defined by 
\begin{align}
\label{eq:max}
\hmax{}(A|B)_\rho\equiv\max_{\sigma^B}\,\, 2\log F(\rho^{AB},\mathbbm{1}^A\otimes\sigma^B),
\end{align}
where the maximization is over positive, normalized states $\sigma$ and $F(\rho,\sigma)\equiv\|\sqrt{\rho}\sqrt{\sigma}\|_1$ is the fidelity of $\rho$ and $\sigma$. Dual to the conditional max-entropy is the conditional min-entropy,
\begin{align}
\hmin{}(A|B)_\rho &\equiv \max_{\sigma^B}\left(-\log \lambda_{\min}(\rho^{AB},\sigma^{B})\right),
\end{align}
with $\lambda_{\min}(\rho^{AB},\sigma^B){\equiv}\min\left\{\lambda:\rho^{AB}\leq \lambda\mathbbm{1}^A\otimes \sigma^B\right\}$. The two are dual in the sense that
\begin{align} \label{eq:dual}
  \hmax{}(A|B)_{\rho}=-\hmin{}(A|C)_{\rho}
\end{align}
for $\rho^{ABC}$ a pure state~\cite{konig_operational_2009}. The min- and max-entropies derive their names in part from the following relation (Lemma 2,~\cite{tomamichel_fully_2009}),
\begin{align}
\label{eq:entropyorder}
\hmin{}(A|B)_\rho\leq H(A|B)_\rho\leq \hmax{}(A|B)_\rho,
\end{align}
where $H(A|B)_\rho\equiv H(AB)_\rho-H(B)_\rho$ for $H(B)_\rho=-{\rm Tr}[\rho\log\rho]$ is the usual conditional von Neumann entropy. 
 
The min- and max-entropies can be \emph{smoothed} by considering possibly subnormalized states $\bar{\rho}^{AB}$ in the $\epsilon$-neighborhood of $\rho^{AB}$, defined using the purification distance 
$P(\rho,\sigma)\equiv\sqrt{1-F(\rho,\sigma)^2}$,
\begin{align}
B_\epsilon(\rho)\equiv\{\bar{\rho}:P(\rho,\bar{\rho})\leq \epsilon\}.
\end{align}  
Note that the purification distance is essentially equivalent to the trace distance, due to the bounds 
$D(\rho,\sigma)\leq P(\rho,\sigma)\leq\sqrt{2D(\rho,\sigma)}$~\cite{tomamichel_duality_2009}.
The smoothed entropies are then given by 
\begin{align}
\hmin{\epsilon}(A|B)_\rho&\equiv\max_{\bar{\rho}\in B_\epsilon(\rho^{AB})} \hmin{}(A|B)_{\bar{\rho}},\\
\label{eq:hmaxsmooth}
\hmax{\epsilon}(A|B)_\rho&\equiv\min_{\bar{\rho}\in B_\epsilon(\rho^{AB})} \hmax{}(A|B)_{\bar{\rho}}.
\end{align}
Furthermore, the dual of $\hmax{\epsilon}(A|B)_\rho$ is $\hmin{\epsilon}(A|C)_\rho$, so that taking the dual and smoothing can be performed in either order~\cite{tomamichel_duality_2009}. 
 
\section{One-shot privacy amplification and data compression with side information} Optimal one-shot privacy amplification results are established in~\cite{renner_security_2005,renner_universally_2005,koenig_sampling_2007,tomamichel_leftover_2010}. Using the entropy definitions above, the number of $\epsilon$-good random bits $\lext{\epsilon}(X|E)$ which can be extracted from the classical random variable $X$ against a possibly quantum eavesdropper is bounded by \begin{theorem}[Privacy Amplification] Given a state $\psi^{XE}=\sum_x p_x \ketbra{x}^X\otimes \varphi_x^E$ and $\epsilon_1,\epsilon_2\geq 0$ such that $\epsilon=\epsilon_1{+}\epsilon_2$, \label{thm:pa} \begin{align*} \hmin{\epsilon_1}(X|E)_\psi-2\log\tfrac{1}{\epsilon_2}\!+\!1\leq \lext{\epsilon}(X|E)_\psi\leq \hmin{\sqrt{2\epsilon}}(X|E)_\psi.  \end{align*} \end{theorem}

Meanwhile, optimal one-shot information reconciliation results are given in~\cite{renes_one-shot_2010}. The minimum number of bits $\lenc{\epsilon}(X|B)$ to which the classical random variable $X$ can be compressed and still be recovered using side information $B$ at the decoder with error probability less than $\epsilon$ is bounded by
\begin{theorem}[Classical-Quantum Information Reconciliation]
\label{thm:csi}
Given a state $\psi^{XB}=\sum_x p_x \ketbra{x}^X\otimes \varphi_x^B$ and $\epsilon_1,\epsilon_2\geq 0$ such that $\epsilon=\epsilon_1{+}\epsilon_2$
\begin{align*}
\hmax{\sqrt{2\epsilon}}(X|B)_\psi\leq \lenc{\epsilon}(X|B)_\psi&\leq \hmax{{\epsilon_1}}(X|B)_\psi+2\log\tfrac{1}{\epsilon_2}+4.
\end{align*}
\end{theorem}

\section{Miscellaneous Facts}

\begin{lemma}[Preimage Sizes of Linear Functions]
\label{lem:linearpreimage}
Let $f:\mathcal{X}\rightarrow \mathcal{Y}$ be a linear function. Then $|f^{-1}(y)|=|\mathcal{X}|/|\mathcal{Y}|$ for all $y\in \mathcal{Y}$.
\end{lemma}
\begin{IEEEproof}
Pick an $x^*$ and consider all the ${x}_j\in \mathcal{X}$ such that $f({x}_j)=f({x}^*)$. Forming the differences $w_j={x}_j-x^*$, it follows from linearity that $f(w_j)=0$ for all $j$. Now consider an arbitrary $x'\in \mathcal{X}$. Clearly $f(x'+w_j)=f(x')$ for all $j$, so each output value has the same number of preimages.
\end{IEEEproof}

\begin{lemma}[Preimage Sizes of Arbitrary Functions]
\label{lem:arbpreimage}
Let $f:\mathcal{X}\rightarrow \mathcal{Y}$ be an arbitrary function and denote by $\mathcal{X}_y$ the preimage of an output $y$. For a randomly-chosen output value $y$, $|\mathcal{X}_y|\geq \eps |\mathcal{X}|/|\mathcal{Y}|$ with probability at least $1-\eps$. 
\end{lemma} 
\begin{IEEEproof}
Let $X$ be a uniform random variable over $\mathcal{X}$ and $Y=f(X)$. By the min-entropy chain rule~(Lemma 3.1.10 in~\cite{renner_security_2005}) we have 
\begin{align}\hmin{}(X|Y)&\geq \hmin{}(XY)-\log\left|{\rm supp}(P_Y)\right|\nonumber\\
&=\hmin{}(X)-\log\left|{\rm supp}(P_Y)\right|\geq\log |\mathcal{X}|/|\mathcal{Y}|.
\end{align}
Here $|{\rm supp}(P_Y)|$ is the size of the support of the distribution $P_Y$, the number of values taking nonzero probability. 
By the normalization condition for $P_{X|Y=y}$ it follows that $1/|\mathcal{X}_y|\leq \max_x P_{X|Y=y}(x)$. And by the definition of $\hmin{}(X|Y)$,
\begin{align}
\sum_{y\in\mathcal{Y}}P_{Y=y}\frac{1}{|\mathcal{X}_y|}\leq 2^{-\hmin{}(X|Y)}\leq \frac{|\mathcal{Y}|}{|\mathcal{X}|}.
\end{align}
Finally, applying the Markov inequality to the random variable $|\mathcal{X}_Y|$ yields 
\begin{align}
{\rm Pr}\left[\frac{1}{|\mathcal{X}_y|}\geq \frac{|\mathcal{Y}|}{\eps|\mathcal{X}|}\right]\leq \frac{\eps|\mathcal{X}|}{|\mathcal{Y}|}\sum_{y\in\mathcal{Y}}P_{Y=y}\frac{1}{|\mathcal{X}_y|}\leq \eps,
\end{align}
which concludes the proof.
\end{IEEEproof}

\begin{lemma}[Smooth Entropy Bounds]
\label{lem:entropybounds}
\begin{align}
\hmin{\eps}(A|B)_\rho&\leq H(A|B)_\rho+8\eps\log {\rm dim}(A)+2h_2(2\eps)\\
\hmax{\eps}(A|B)_\rho&\geq H(A|B)_\rho-8\eps\log {\rm dim}(A)-2h_2(2\eps),
\end{align}
where $h_2(x)=-x\log_2 x-(1-x)\log_2(1-x)$ is the binary entropy function. 
\end{lemma}
\begin{IEEEproof}
Let $\bar{\rho}$ be a state in 
$B_\epsilon(\rho)$ such that $\hmin{}(A|B)_{\bar{\rho}}=\hmin{\eps}(A|B)_\rho$. Then from Eq.~\ref{eq:entropyorder} we have $\hmin{\eps}(A|B)_\rho\leq H(A|B)_{\bar{\rho}}$. Since the purification distance bounds the trace distance, $D(\rho,\bar{\rho})\leq \eps$ and we can use the continuity of the conditional von Neumann entropy~\cite{alicki_continuity_2004} to establish that 
$H(A|B)_{\bar{\rho}}\leq H(A|B)_\rho+8\eps\log {\rm dim}(A)+2h_2(2\eps)$, completing the proof for the min-entropy. An entirely similar argument holds for the max-entropy.
\end{IEEEproof}

\begin{lemma}[Single-Letter Formula for the Private Capacity]
\label{lem:singlelett}
For a channel $\Theta_{\rm cl}$ with purely classical outputs $Y$ and $Z$, 
\begin{align}
R_\prv(\Theta_{\rm cl})&\leq\max_{P_T,T\rightarrow X}\left[H(T|Z)-H(T|Y)\right].
\end{align}
\end{lemma}
\begin{IEEEproof}
Start with the expression $b=H(T|Z^{\otimes \ell})-H(T|Y^{\otimes \ell})$ from Eq.~\ref{eq:privrate}. By Lemma 4.1 of~\cite{ahlswede_common_1993}, or direct calculation, we can rewrite this as a sum
\begin{align} 
b=\sum_{i=1}^\ell b_i\equiv \sum_{i=1}^\ell H(T|V_iZ_i)-H(T|V_iY_i)
\end{align}
for $V_i=Z_1\dots Z_{i-1}Y_{i+1}\dots Y_\ell$. Observe that the random variables involved in $b_i$ 
form the Markov chain $(T,V_i)\leftrightarrow X_i\leftrightarrow (Y_i,Z_i)$. This follows because $(X_i,X',Y',Z')\leftrightarrow X_i\leftrightarrow (Y_i,Z_i)$ is a Markov chain, where $X'$ denotes the tuple of $X_j$ random variables omitting $X_i$, and $(T,V_i)$ can be computed from $(X_i,X',Y',Z')$. 

But now we can maximize each term over $(T,V_i)$ subject to the Markov chain condition and obtain the single-letter formula
\begin{align*}
b&\leq \sum_{i=1}^\ell \max_{(T,V_i)}\left[H(T|V_iZ_i)-H(T|V_iY_i)\right]\\
&\leq  \max_{(T,V)}\ell\left[H(T|VZ)-H(T|VY)\right],
\end{align*}
for the Markov chain $(T,V)\leftrightarrow X\leftrightarrow (Y,Z)$.
Since Shannon entropies conditioned on $V$ are averages of entropies conditioned on specific values $V=v$, this implies
\begin{align*}
b&\leq  \max_{(T,V,v)}\ell\left[H(T|Z,V=v)-H(T|Y,V=v)\right].
\end{align*}
Finally, conditioning on $V=v$ preserves the Markov chain $T\leftrightarrow X\leftrightarrow (Y,Z)$, since $P_{Y Z | X} = P_{Y Z | X T V}$ implies $P_{Y Z | X, V=v} = P_{Y Z | X T, V=v}$. And because each choice of $V$ and $v$ induces a conditional distribution $P_{T|V=v}$, the maximization need only be taken over $T$. Using this in Eq.~\ref{eq:privrate} completes the proof. 
\end{IEEEproof}

\end{document}